\documentclass[11pt]{article}
\usepackage{hyperref}
\pdfoutput=1
    \usepackage{graphicx}
\usepackage{color}
\usepackage{subfig}
\usepackage{lscape}
\usepackage[numbers,square,sort&compress]{natbib}
\usepackage{amsmath}
\usepackage{amssymb}
\usepackage{wrapfig}
\usepackage{hyperref}
 \pdfoutput=1

\begin{document}

\title{Dynamics of Ferrofluidic Drops Impacting Superhydrophobic Surfaces}
\author{D.A. Bolleddula, H.E. Dillon, A. Alliseda, P. Bhosale, J.C. Berg  \\
\\\vspace{6pt} Mechanical and Chemical
Engineering, \\ University of Washington}
\maketitle
\begin{abstract}
This is a fluid dynamics video illustrating the impact of
ferrofluidic droplets on surfaces of variable wettability.
Surfaces studied include mica, teflon, and superhydrophobic. A magnet is placed beneath each surface, which modifies the behavior of the ferrofluid by applying additional downward force apart from gravity resulting in reduced droplet size and increased droplet velocity. For the superhydrophobic droplet
a jetting phenomena is shown which only occurs in a limited range of impact speeds, higher than observed before, followed by amplified oscillation due to magnetic field as the drop stabilizes on the surface.
\end{abstract}
\section{Introduction}
Ferrofluids have attracted attention in microfluidics as a novel
method for a passive transport mechanism. Recent studies \cite{ferro1} have shown water based ferrofluid droplets can be driven magnetically on a superhydrophobic surfaces and liquid marbles (droplet encapsulated in superhydrophobic powder coating) on hydrophilic surfaces. Here the effects of surface wetting and magnetic actuation on the dynamics of droplet impact are illustrated briefly in the video as an introduction to the fluid dynamics of free surface flows. 

This video explores the impact of ferrofluid droplets and liquid
marbles on several surfaces including hydrophilic mica, hydrophobic teflon, and superhydrophobic treated fumed silica coating. The superhydrophobic surfaces were prepared by dip coating treated fumed silica powder (Cab-O-Sil TS530, Cabot Corp.) through acetone solution on a glass slide with a thin layer of epoxy resin (5 Minute Epoxy, ITW Devcon Corp.). The liquid marbles were formed by coating droplets with TS 530 powder. The magnet is placed below the surface and the video was recorded with a high speed camera system.
\href{URL of video}{Video}.

\begin{figure}[h]
  \centering
    \subfloat[mica]{\includegraphics[trim = 0mm 0mm 0mm 0mm, clip, width=0.6\textwidth]{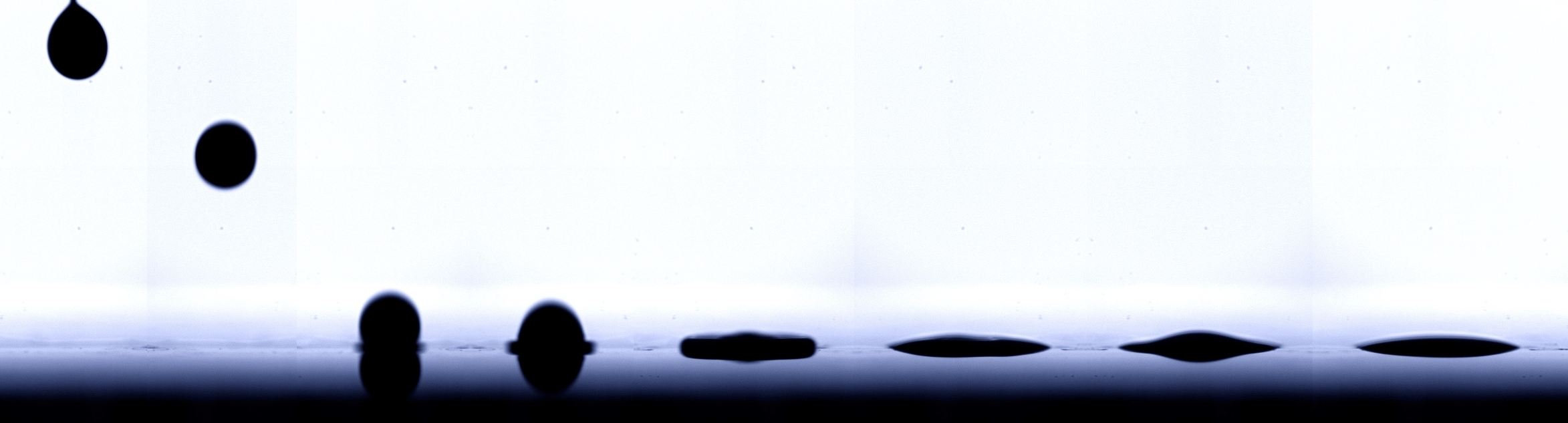}}

    \subfloat[teflon]{\includegraphics[trim = 0mm 0mm 0mm 0mm, clip, width=0.7\textwidth]{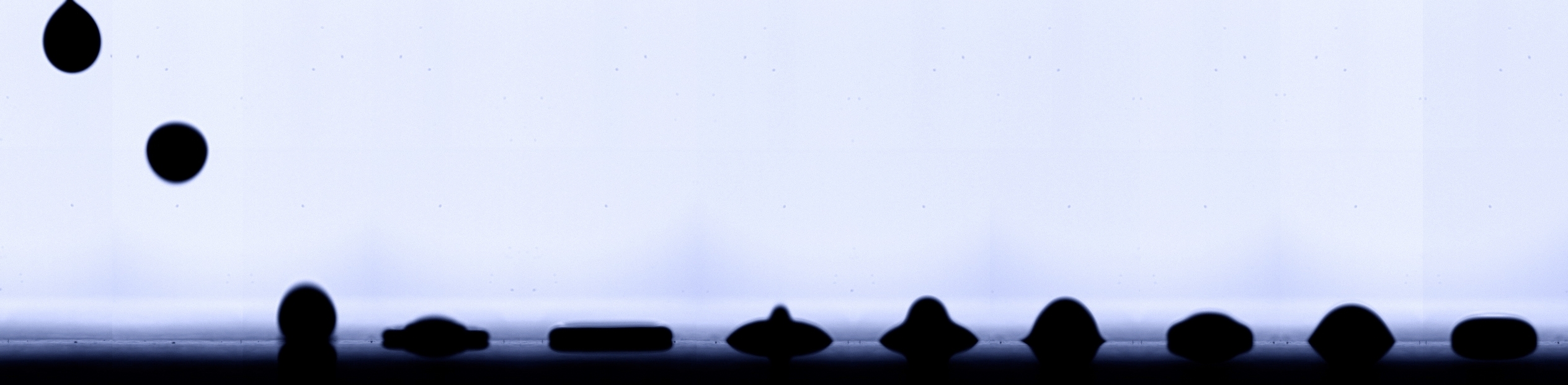}}

    \subfloat[superhydrophobic]{\includegraphics[trim = 0mm 0mm 0mm 0mm, clip, width=0.99\textwidth]{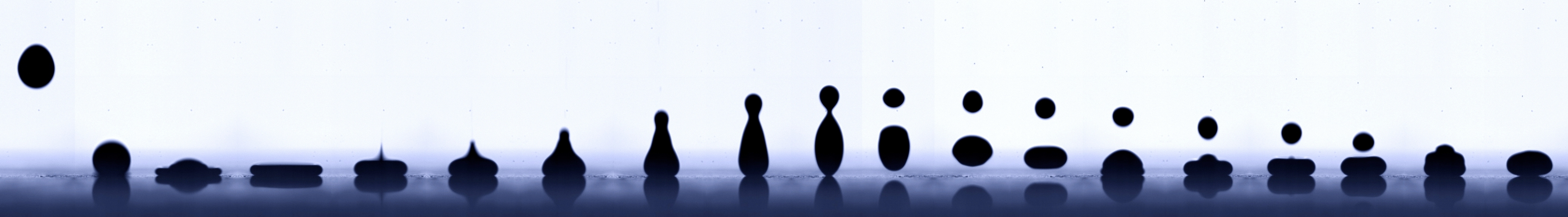}}
  \caption{Impact of a ferrofluidic droplet on mica, teflon, and a superhydrophobic surface.}
  \label{fig:impact}
\end{figure}

A summary of the surface impacts are shown in Figure
\ref{fig:impact}. On the mica surface (a), the droplet impacts the
surface and spreads to equilibrium due to the high surface energy.
On the teflon surface (b), the droplet impacts, then spreads and
recedes with several periods of damped oscillations amplified by
the magnetic field strength. On the superhydrophobic surface (c)
the drop impacts and creates an air cavity. The air cavity
collapses and causes a jet to eject a small drop at 6 times the
initial impact velocity. 

\newpage

\begin{figure}[h]
  \centering
   \includegraphics[trim = 0mm 0mm 0mm 0mm, clip, width=0.4\textwidth]{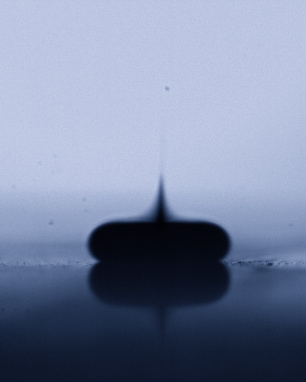}
  \caption{High velocity jet leaving ferrofluid after impact on a superhydrophobic surface.}
  \label{fig:jet}
\end{figure}

The jetting behavior is known to occur
only in a narrow range of impact speeds ($We\sim$ 8-40) which is a
larger range observed by \cite{Bartolo06}. In this case, the jet
phenomena is caused by the increased velocity induced by the
magnet and ferrofluid. Figure \ref{fig:jet} shows an image from
the superhydrophobic impact where the high velocity ejected drops
are shown more clearly.


\bibliographystyle{unsrtnat}
\bibliography{ferrodrops.bib}

\end{document}